\documentstyle[12pt,aaspp4]{article}

\newcommand{\etal}{{\it{et al.}}~}

\def\ltaprx{\, \buildrel < \over \sim \,}

\begin{document}

\title{Deep HST Observations of Star Clusters in NGC 1275
\footnote{Based
on observations with the NASA/ESA {\it Hubble Space Telescope},
obtained at
the Space Telescope Science Institute, operated by AURA Inc under
contract
to NASA}}

\author{
Matthew~N.~Carlson\altaffilmark{2},
Jon~A.~Holtzman\altaffilmark{2},
Alan~M.~Watson\altaffilmark{3},
Carl~J.~Grillmair\altaffilmark{4},
Jeremy~R.~Mould\altaffilmark{5},
Gilda~E.~Ballester\altaffilmark{6},
Christopher~J.~Burrows\altaffilmark{7},\\
John~T.~Clarke\altaffilmark{6},
David~Crisp\altaffilmark{4},
Robin~W.~Evans\altaffilmark{4},
John~S.~Gallagher~III\altaffilmark{8},\\
Richard~E.~Griffiths\altaffilmark{9},
J.~Jeff~Hester\altaffilmark{10},\\
John~G.~Hoessel\altaffilmark{8},
Paul~A.~Scowen\altaffilmark{10},
Karl~R.~Stapelfeldt\altaffilmark{4},
John~T.~Trauger\altaffilmark{4},
and
James~A.~Westphal\altaffilmark{11}
}

\altaffiltext{2}{Department of Astronomy, New Mexico State
University, Dept 4500
Box 30001, Las Cruces, NM 88003, mcarlson@nmsu.edu, holtz@nmsu.edu}
\altaffiltext{3}{Instituto de Astronom\'\i a UNAM, J. J. Tablada
1006, Col. Lomas de Santa Maria, 58090 Morelia, Michoacan,
Mexico, alan@astrosmo.unam.mx}
\altaffiltext{4}{Jet Propulsion Laboratory, 4800 Oak Grove Drive,
Pasadena,
CA 91109, carl@wfpc2-mail.jpl.nasa.gov, dc@crispy.jpl.nasa.gov,
rwe@wfpc2-mail.jpl.nasa.gov, krs@wfpc2-mail.jpl.nasa.gov,
jtt@wfpc2-mail.jpl.nasa.gov}
\altaffiltext{5}{Mount Stromlo and Siding Spring Observatories,
Australian National University, Private Bag, Weston Creek Post
Office, ACT 2611, Australia,
jrm@mso.anu.edu.au}
\altaffiltext{6}{Department of Atmospheric, Oceanic, and Space Sciences,
University of Michigan, 2455 Hayward, Ann Arbor, MI 48109,
gilda@sunshine.sprl.umich.edu, clarke@sunshine.sprl.umich.edu}
\altaffiltext{7}{Astrophysics Division, Space Science Department,
ESA \& Space
Telescope Science Institute, 3700 San Martin Drive, Baltimore, MD
21218,
burrows@stsci.edu}
\altaffiltext{8}{Department of Astronomy, University of Wisconsin
-- Madison, 475 N. Charter St., Madison, WI 53706,
jsg@tiger.astro.wisc.edu,hoessel@jth.
astro.wisc.edu}
\altaffiltext{9}{Department of Physics, Carnegie Mellon
University, 5000 Forbes
Ave, Pittsburgh, PA 15213}
\altaffiltext{10}{Department of Physics and Astronomy, Arizona
State University,
Tyler Mall, Tempe, AZ 85287, jjh@cosmos.la.asu.edu,
scowen@tycho.la.asu.edu}
\altaffiltext{11}{Division of Geological and Planetary Sciences,
California Institute of Technology, Pasadena, CA 91125,
jaw@sol1.gps.caltech.edu}

\begin{abstract}

We present an analysis of compact star clusters in deep HST/WFPC2 images of NGC
1275.  B and R band photometry of roughly 3000 clusters shows a
bimodality in the B-R colors, suggesting that distinct old and young
cluster populations are present.  The small spread in the colors of the
blue clusters is consistent with the hypothesis that they are a single
age population, with an inferred age of 0.1 to 1 Gyr.
The luminosity function shows increasing numbers of
blue clusters to the limit of our photometry, which reaches several
magnitudes past the turnover predicted if the cluster population were
identical to current Galactic globulars seen at a younger age. The blue
clusters have a spatial distribution which is more centrally peaked
than that of the red clusters.
The individual clusters are slightly
resolved, with core radii $\ltaprx 0.75$ pc if they have modified
Hubble profiles.
We estimate the specific frequencies of the old and young populations
and discuss the uncertainties in these estimates.
We find that the specific frequency of the young population in NGC 1275
is currently larger than that of the old population and will remain so
as the young population evolves, even if the majority of the low mass
clusters are eventually destroyed. If the young population formed during a
previous merger, this suggests that mergers can increase the specific
frequency of globulars in a galaxy. However, the presently observed young
population likely contains too few clusters to have a significant impact
on the overall specific frequency as it will be observed in the future.

\end{abstract}

\section{Introduction}

Early HST observations of NGC 1275, the central galaxy in the Perseus
cluster, revealed a population of about
60 blue (V-R $\sim$ 0.3) star clusters surrounding the nucleus 
(Holtzman \etal \markcite{hol92} 1992). The color
of these clusters suggests an age of roughly 300 million years based
on the models of Charlot and Bruzual \markcite{cb91} (1991).
Spectra of the brightest
object (Zepf \etal \markcite{zep95} 1995) also suggest an age of 0.1
to 0.9 Gyr, based on a
comparison of the observed line widths to
those predicted by the models of Bruzual and Charlot \markcite{bc93} (1993).
None of the clusters seem to have H$\alpha$
emission, with the exception of one object found by Shields \&
Filippenko \markcite{sf90} (1990), which appears to be a much younger object.
The blue
clusters in the original WFPC1 images appear unresolved, suggesting sizes of
less than 15 parsecs.  The brightest object has V = 18.9, which
corresponds to M$_V$ = -15.8
for $H_0$ = 75 km/s/Mpc and cz=5264 km/s
(Strauss \etal \markcite{str92} 1992). The brightnesses of the blue
clusters suggest masses
of between 2 $\times$
10$^4$ and 1 $\times$ 10$^8$ M$_{\odot}$, depending on the assumed age
and distance, if one assumes that the objects are star clusters with a Salpeter
IMF. The observed sizes, luminosities, and the estimated
masses suggest that these objects may be young analogues of globular clusters.

In the past several years, massive young clusters have been observed in
a variety of other galaxies.  Lutz \markcite{lut91} (1991) detected
young globular cluster candidates
in a ground-based study of the merger remnant
NGC 3597, and these have been confirmed to be compact by HST
observations (Holtzman \etal \markcite{hol96} 1996).  Candidate young globular clusters
have also been found in other interacting systems, including NGC 7252
(Whitmore \etal \markcite{whit93} 1993), NGC 4038/9
(Whitmore \& Schweizer \markcite{ws95} 1995), NGC
3921 (Schweizer \etal \markcite{sch96} 1996), among others.
A few massive clusters are seen in the
starburst galaxy NGC 253 (Watson \etal \markcite{wat96} 1996), while others have been
detected in the ring galaxies NGC 1097 and NGC 6951 (Barth \etal
\markcite{bhfs95} 1995) and the dwarf galaxies NGC 1569, NGC 1705
(O'Connell \etal \markcite{ogh94} 1994),
and He 2-10 (Conti \& Vacca \markcite{cv94} 1994).
While
no generally accepted picture has emerged as to what conditions lead to
the formation of young clusters,
galaxy interactions appear to be an important
component.

It is particularly
difficult to determine the precise mechanism responsible for
the presence of these
objects in NGC 1275 because of the myriad peculiarities
of this galaxy, including the presence of a significant amount of
dust, streamers of H$\alpha$ emission, an active nucleus, and the location
of the galaxy at the center of a cooling flow in the Perseus cluster.
Two hypotheses have been offered for the origin of these clusters,
namely, that they were formed from the substantial mass deposition of
the Perseus cluster cooling flow (200 M$_{\odot}$/yr)
or that the cluster formation was
triggered by a galaxy-galaxy interaction.  Holtzman \etal \markcite{hol92} (1992) preferred
the latter hypothesis based on the observed lack of spread in the WFPC1
colors, which implies a common age for the objects, and the appearance
of a ripple in the galaxy light which suggests that a previous
interaction may have occurred. Richer \etal \markcite{ric93} (1993)
found
larger color spreads in high resolution CFHT images and preferred the
cooling flow hypothesis. However,
Holtzman \etal \markcite{hol96} (1996) did not
detect young clusters in three of
a sample of four other cooling flow galaxies; the central cluster
galaxy in Abell 1795 may have a few clusters, but it also has a
peculiar morphology which suggests a previous interaction.

The merger hypothesis is particularly interesting in light of the
theory that elliptical galaxies form through mergers and the
observation that ellipticals have higher specific frequencies of
globular clusters than spirals, where the specific frequency is
a measure of the
number of globular clusters per unit luminosity of the
host galaxy. Ashman \& Zepf \markcite{az92} (1992) investigate the
proposition that clusters might form during mergers and suggest that,
in such an event, cluster formation would occur in a brief burst,
resulting in a set of newly formed clusters with a
common age and color.  This would lead to a cluster system with a
bimodal distribution of cluster colors reflecting the difference in
metallicity or age (or both) between the original and the newly formed
clusters. They predict that the spatial distribution of the younger
clusters would be more sharply peaked toward the center of the galaxy
than that of the old cluster system, because the old globular cluster
populations of the two progenitor galaxies would remain spatially
extended and probably be dynamically heated
during the merger, while the new clusters would be formed out
of gas which becomes more centrally concentrated during a merger.
However, as Harris \etal \markcite{har95} (1995) note,
an increase in specific frequency
as a result of a merger requires not only that globular clusters form
during such an event, but that they form {\it preferentially} over 
non-cluster stars as compared to the ratio of clusters to background 
stars in the
progenitor galaxies.

Few estimates of the specific frequency of
these young cluster systems exist. Watson \etal \markcite{wat96}
(1996) suggest
that the 4 young clusters in NGC 253 most likely have formed with a
large specific frequency. In the merger remnant NGC 3921 Schweizer
\etal \markcite{sch96} (1996) find that the
blue clusters will increase the overall number of clusters
enough that the galaxy will come to have the specific
frequency of an elliptical within 7 Gyr. Miller \etal \markcite{mil97}
(1997) find that the specific frequency of globular clusters in
the merger remnant NGC 7252 will rise over the next 15 Gyr
as the background population fades to resemble that of an elliptical.
These results all suggest that interactions can increase the specific 
frequency in a galaxy.

Luminosity functions of old globular cluster systems have been well
studied and used as a part of the extragalactic distance scale
because of their uniformity from galaxy to galaxy; they are roughly
Gaussian in shape with a peak near M$_V$ = -7.3 (Harris \markcite{har96} 1996).
The luminosity
functions of most of the recently discovered young cluster systems are
poorly determined because of either small number statistics (few young
clusters in the galaxy) or incompleteness. A notable exception is the
young cluster system in NGC 4038/9, which shows no turnover to two
magnitudes fainter than the turnover predicted for a typical old
cluster system, even after allowing for the expected fading of the
clusters based on stellar population models.  Based on this, van den
Bergh \markcite{vdb95} (1995) argues that this cluster system may
be intrinsically different from old globular cluster systems.
However, Mateo \markcite{ma93} (1993) notes that, at least
in the LMC, the observed increasing cluster luminosity function
is well modelled by a combination of globular and
open clusters. Also, several authors
have recently suggested that substantial numbers of clusters
may be destroyed over a Hubble time (Gnedin \&
Ostriker \markcite{go97} 1997;
Elmegreen \& Efremov \markcite{ee97} 1997);
mechanisms include evaporation and tidal
disruption by galactic bulges and disks.

We have obtained WFPC2 images of NGC 1275 which go about 4 magnitudes 
deeper (in the red) than the WFPC1 observations
of Holtzman et al. \markcite{hol92} (1992).
These observations provide more accurate colors
than previous observations and probe the cluster population
to approximately 2.5 magnitudes fainter than
the turnover expected if this population is identical to the Galactic
globular cluster system seen at an younger age.

Section 2 briefly summarizes the observations and the reductions. Section 3
discusses our analysis and potential sources of error. In section 4 we
present our results including the photometry, luminosity function of the
clusters, surface density distribution, an estimate of the specific frequency,
and analysis of the sizes of the objects.

\section{Observations \& Data Reduction}

Observations of NGC 1275 were made with the WFPC2 on 1995 November 16
in the F450W and F702W filters. Exposure times in F450W
were 200, 1000, and 3 $\times$ 1300 seconds.  In F702W the
exposures were 200, 900, 1000, and 2 $\times$ 1300 seconds. The F450W and
F702W filters are roughly similar to broad band B and R filters.

The data were reduced using the procedures discussed by Holtzman \etal
\markcite{hol95a} (1995a). The individual exposures were all taken with a common pointing,
and the frames in each color were combined. Cosmic rays were rejected
in the averaging based on the expected variance from photon statistics
and readout noise; an extra term was included in the expected variance which
was proportional to the signal in the pixel to allow for small pointing
differences between frames.

\section{Analysis}

Combined color images of NGC 1275 in the PC and in the WFPC2
fields are shown in Figures 1a and 1b, respectively.  The nucleus is in
the center of the PC, and the other bright point source in the PC is a
foreground F star (Hughes \& Robson \markcite{hr91} 1991).  Almost all of the remaining
point sources are candidate star clusters in NGC 1275.

To detect the compact clusters, we summed the two colors and
subtracted a 5 $\times$ 5 boxcar-smoothed
version of this image.
We divided this by the square root of the smoothed image to provide an
image whose units were proportional to the local noise (assuming the
noise is dominated by the background, as is the case here). We then used
IRAF's DAOFIND to detect objects on the
resulting image, in order to provide a uniform detection threshold in
signal-to-noise ratio across the image.  The choice of detection threshold was
determined by minimizing the number of spurious detections and
maximizing the number of faint objects detected, judging from a visual
inspection.  Objects were filtered by shape based on
roundness and sharpness criteria computed by DAOFIND; we required
-1 $<$ \it{roundness}\rm \ $<$ 1 and 0.2 $<$ \it{sharpness}\rm \ $<$ 1.
Several areas of the frame were masked out, including the galaxy
nucleus, the bright saturated star in the PC, and a few bright stars
and galaxies in the WFs.

Aperture photometry was performed using a 1-pixel radius aperture in
both the PC and the WFs. The small aperture was used to reduce
background related errors, which are large in the central
region of the galaxy.  Sky values were determined by taking an estimate
of the mode in
a background annulus; in the PC, the annulus was 2 pixels wide with an
inner radius of 12 pixels, while in each WF the corresponding annulus
was 3 pixels wide with an inner radius of 6 pixels. We chose to use 
small sky annuli because the galaxy background is variable on small scales.

Corrections from
1-pixel aperture flux to flux within 0.5$^{\prime \prime}$ were
determined from the nine brightest sources.  The average aperture
corrections were 1.03 and 1.15 magnitudes in the PC frame for F450W and
F702W, respectively, and 0.63 and 0.71 magnitudes in the WFs.  These
are slightly larger than the corrections for bright point sources
measured by Holtzman \etal \markcite{hola95} (1995a), but this is expected if these
objects are slightly resolved. In the PC, measured aperture corrections
varied on the order of 0.05 mag between different objects.
Measured variations in the aperture
corrections for the individual WFs were $\sim$ 0.02 mag, so an average
aperture correction was assumed for all three.
If there is an intrinsic spread in cluster sizes, using a single aperture
correction for all objects would lead to systematic errors in the derived
magnitudes. To first order, however, colors
are unaffected by such a spread.
The transformations
from HST magnitudes (F450W, F702W) to Cousins magnitudes (B, R) were
made using the transformations given by Holtzman \etal \markcite{holb95} (1995b).

Foreground Galactic extinction in the direction of NGC 1275 was taken
to be $A_B$ = 0.70 (Burstein \& Heiles \markcite{bh84} 1984).  We determined the
extinction in the WFPC2 passbands by numerically integrating the
product of an A star spectrum, the filter transmission, the system response,
and an
extinction curve from Cardelli \etal \markcite{ccm89} (1989) derived assuming $R_V=3.1$.
An A star was used because of its similarity to the observed spectrum
of the bright clusters. This gave $A_{F450W} = 0.69$ and $A_{F702W} =
0.40$.  These estimates do not include internal extinction, however,
which may be important in NGC 1275 since patchy dust is evident;
this is discussed further below.

Potential photometry errors include intrinsic Poisson error in the
signal, error in the aperture correction, and error in the background
level determination. Typical errors were estimated from the simulated
clusters from our completeness tests and are shown in Figure 2.
The dashed lines in color and
magnitude indicate input magnitudes for the artificial clusters.

Completeness was estimated by generating simulated objects at each of
9 different
input magnitudes (23.75 to 27.75 in B). The simulated objects were given 
the same noise characteristics as real objects.
In the PC, 1000 objects were placed randomly in each of 3 separate annuli to
better model the variation of completeness with distance from the
center of the galaxy.
Each set of artificial clusters was
given a spatial distribution which mimicked the distribution
of the brightest (B $<$ 25) blue and red clusters.
In the WFs, the 1000
artificial objects were placed randomly in a uniform distribution.
On each chip, a PSF of one of the 
isolated bright clusters was used to create the simulated objects (the use
of other objects gave comparable results).  
Two sets of artificial objects were
placed to estimate the difference in detection efficiency for objects
of two different colors; the two sets of simulated clusters were given
B-R=0.7 and B-R=1.6, comparable to the colors of the observed
cluster populations (next section).
The derived average
completeness (weighted by the spatial distribution of the observed
clusters) as a function of B magnitude is shown in Figure 3 for the
whole image as well as for the section which fell in the PC.

We estimated the number of foreground stars based on the models of
Bahcall and Soneira \markcite{bs80} (1980) and expect $<$ 4 objects with B
$<$ 27 in the PC.  Since we expect such small numbers
of foreground stars, we have neglected them.

\section{Results}

\subsection{Photometry}

We present the photometry of all detected objects in NGC 1275 by
plotting color against apparent magnitude in Figure 4a. 
Extinction-corrected absolute magnitudes and colors are shown as well, using a
distance of 70.2 Mpc
determined by assuming pure Hubble flow with $H_0$ = 75
km/s/Mpc.  
We find 1181 blue objects (B-R $<$ 1.1) and 1855 red objects.  
Distinct red and blue populations are seen, although there is significant
color scatter. Roughly 50\% of the blue objects are seen in the PC,
as compared to about 15\% of the red objects.  

These objects cannot be single stars because they are marginally resolved
and because they are too bright. With the exception of
the Shields-Filippenko object, the objects
lack H$\alpha$ emission (Holtzman \etal \markcite{hol92} 1992), which precludes their being
emission line objects. Consequently, they are probably star clusters, as
supported by spectroscopy of the brightest
object (Zepf \etal \markcite{zep95} 1995).
They are brighter and more compact than open clusters in the Galaxy; the
brightest object is 250 times brighter than the brightest
Galactic open cluster.
Their closest analog seems to be globular clusters,
although the inferred ages are much smaller (see below).

To test whether the observed color scatter is likely to be related to internal
extinction within NGC 1275 and photometric errors from errors in background
subtraction, we show  in Figure 4b
only clusters in the southwestern half of
the PC frame which have estimated photometric errors of less than 0.15 mag in 
both B and R.
This region was chosen because it
contains fewer obvious signatures of dust than the northeastern region.  
In fact, we find that the spread in color among the blue objects in Figure 4b 
is consistent with that associated with photometric errors (cf. Figure 2), 
supporting the conclusion that the observed color spread within the two
populations is likely to be caused by the variation of internal extinction
in the environment of the galaxy, rather than
differences in age.
We conclude that the blue clusters are a single color, single age
population, with the spread in colors coming from Poisson errors and
background subtraction errors.
Figure 4b shows a distinct bimodality in the cluster population, consistent
with the presence of two discrete populations.

The blue clusters have a typical dereddened (B-R)$_0 \sim$ 0.4.
This is much bluer than typical old globular clusters, which have (B-R)$_0$
$\sim$ 1.2 (Harris \markcite{har96} 1996).  
The uniformity of the colors suggests that the clusters
were formed in a relatively short duration single burst rather than an
ongoing or continuous process. Figure 5 shows the magnitude and color
evolution of a single burst population for a variety of metallicities, 
assuming a Salpeter initial mass function
from 0.1 to 125 M$_\odot$ and a total mass of 1 M$_\odot$ 
(Bruzual \& Charlot \markcite{bc93} 1993).  If we compare the measured
colors of the blue clusters in the PC to the models, we infer ages
of 10$^7$ to 10$^9$ years, allowing for small possible systematic errors
in the models and the data as well as the unknown metallicity of the
objects. We note that the clusters are far too blue to be explained
by a metallicity effect alone. Ages younger than 10$^7$ years are
ruled out by the absence of H$\alpha$ emission from the clusters. 
Spectra of the clusters (Zepf \etal \markcite{zep95} 1995) argue
for a several hundred Myr old
population dominated in the blue by A type stars.
It is also simply less probable that we that we should
happen to observe these clusters at a younger age.

The LMC has a number of compact clusters with comparable colors (van den Bergh
\markcite{vdb81} 1981).
These objects are substantially fainter than the brightest objects in NGC
1275, implying that they are less massive. Still,
observations of blue clusters in the LMC, where we can directly measure
ages using the main sequence turnoff in color-magnitude diagrams, give
us a less model dependent idea of the spread in colors that could be
expected from a spread in ages. Figure 6 shows the distribution
of B-V versus age for LMC clusters; ages are from Elson \& Fall
\markcite{ef88} (1988) and Hodge \markcite{hod81} \markcite{hod83} (1981, 1983),
and colors from van den Bergh \markcite{vdb81} (1981).
Between 10$^8$ and 10$^9$ years, we see that the integrated colors of
the clusters cover a range of 0.3 in B-V. This argues against a large
spread in the ages for the NGC 1275 clusters, which show no such
spread in colors.
While our observations were in B and R, not B and V,
the color spread in B-R for the same age range would be 0.8 magnitudes,
assuming a V-R vs B-R relation from the population synthesis models.

Given ages, we can estimate masses for the clusters based on their
luminosities. Age estimates of 10$^8$ and 10$^9$ years give mass
estimates for the brightest object of 2 $\times$ 10$^7$ and
10$^8$ M$_\odot$, respectively.
Blue clusters at our completeness
limit of B $=$ 27 have masses of 2 $\times$ 10$^4$ to
10$^5$ M$_\odot$. These masses are very comparable with those of Milky
Way globulars, which typically have masses between 10$^4$ and
3$\times$ 10$^6$ M$_\odot$, with a handful of clusters with M $<$
10$^4 $M$_{\odot}$ (Mandushev, Spassova, \& Staneva 
\markcite{mss91} 1991, Pryor and Meylan \markcite{pm93} 1993).
Our mass estimates depend on the assumption of a
Salpeter IMF.

Two of the three brightest objects in the southwest half of the PC
(Figure 4b)
have much redder colors than the
rest of the blue clusters (B-R of 0.95 and 1.23).  While these objects
are close to the nucleus, they are bright enough that errors in
determining the galaxy background are not a large source of error.
Possible explanations are reddening by dust,
a younger age (slightly less than 10$^7$ years old)
when the integrated light is dominated by red supergiants
(Leitherer \& Heckman \markcite{lh95}, 1995), or the identification of these
objects with late type
foreground stars. For the redder object, the last
interpretation is supported by our measurement of sizes ($\S 4.5$), since
it appears to be unresolved while almost all other objects appear
marginally resolved.

We suggest that the objects with B-R $\sim$ 1.6 ($\sim$ 1.3 dereddened)
are the old population of globular clusters around NGC 1275. These are
slightly redder than Galactic globular clusters which have an
average (B-R)$_0$ $\sim$
1.2. The brightest (M$_B$ = -10.7) has a magnitude similar to the
brightest globulars seen in other central cluster galaxies.

\subsection{Luminosity Function}

Figure 7a shows the B luminosity function for the blue clusters (B-R
$<$ 1.1), and Figure 7b shows the B luminosity function for the red
clusters (B-R $>$ 1.1).  The dotted lines in each plot show the
results after correction 
for incompleteness.  The luminosity function of the blue
clusters looks distinctly different from that expected for a typical old
globular cluster population at a younger age.  While the Galactic
globular cluster luminosity function is Gaussian in shape with
peak at M$_B$ $\sim$ -6.6 and $\sigma =$ 1.0 mag (Harris \markcite{har96} 1996),
the luminosity function for the clusters in NGC 1275 is more closely
fit by a exponential.  If we were to observe the current Galactic
globular cluster system at an age of 500 million years and at the distance
of NGC 1275 (and the same Galactic extinction), we would
observe a turnover in the luminosity function at B $\sim$
24 ($H_0 =$ 75 km/s/Mpc).  This prediction assumes
4 magnitudes (13 Gyr) of fading as suggested by the models of Bruzual
\& Charlot \markcite{bc93} (1993).  Our
photometry is complete at the 50\% level to 3
magnitudes fainter than this, yet there is no evidence of a turnover in
the luminosity function of the clusters in NGC 1275.

The luminosity function of the blue clusters has the same trend as that
of Galactic open clusters, namely, an increasing
number of clusters at fainter magnitudes with no turnover (van den Bergh
and LaFontaine \markcite{vl84} 1984). However,
the most luminous clusters in NGC
1275 are brighter than any observed Galactic open clusters, despite being
older than the brightest of them.
The compactness as inferred from the HST images is also different
from that of open clusters ($\S 4.5$).
Mateo \markcite{ma93} (1993) finds that an increasing luminosity function for
clusters in the LMC can be produced by a mixed population of true globulars
and open clusters. In NGC 1275, a similar explanation would require a large
number of open clusters which are far
more massive than any Galactic open clusters.
Consequently, as a system, these clusters are
different from any cluster system
seen in Local Group galaxies. As individual objects, however, the
physical properties appear most similar to globular clusters.

In Figure 8,
we compare the luminosity function of clusters in NGC 1275 with the
luminosity functions in several of the other galaxies where
reasonable numbers of blue clusters are found, as well as with the
luminosity functions of Milky Way clusters.
Dashed lines show rough completeness limits for the different observations,
and dotted lines give the location of the expected turnover given the
various age estimates of the clusters. There is no strong evidence in any of
the young cluster systems  for an intrinsic turnover in the luminosity
function.

If these objects will evolve to look like the Galactic globular
cluster system, some mechanism must preferentially destroy low
luminosity, low mass clusters on timescales of 13 Gyr.  Two
possibilities are tidal disruption and evaporation (Gnedin \& Ostriker
\markcite{go97} 1997; Elmegreen \& Efremov \markcite{ee97} 1997).
For tidal disruption to preferentially destroy low mass clusters,
such clusters must have a lower mean
density than more massive clusters. 
Evaporation through two-body relaxation can
preferentially destroy lower mass clusters as their stars escape the
cluster potential well and are captured by the gravitational potential
well of the galaxy. Gnedin and Ostriker \markcite{go97}
(1997) predict that from half
to three quarters of the initial globular cluster population in a
galaxy may be destroyed by a combination of these processes in a
Hubble time.

\subsection{Surface Density Profile}

Figure 9 shows the surface
density distribution of red and blue clusters (open
triangles and filled triangles, respectively).  Corrections have been
made for incomplete annuli at large radii, for masked areas, and
for differing completeness levels with distance from the bright galaxy
center. Error bars are simple Poisson errors in counts of clusters. The
central area, within 18 pixels (0.8 arcsec) from the core of the
galaxy, has been omitted from the surface density profile due to the
bright nucleus. It is clear that the blue clusters are more centrally 
concentrated than the red ones.

We fit the surface density distribution of the blue and red cluster
systems from 4$^{\prime \prime}$
to 120$^{\prime \prime}$
with power laws. For the blue clusters, we find that a power law slope of
-1.3 provides a reasonable fit over the entire range. For the red clusters,
we note a flattening in the inner portion of the distribution and fit
the surface density with two power laws. Inside 0$^{\prime} \!$.5,
we measure a slope of -0.6, while outside 0$^{\prime} \!$.5
we find a slope of -1.0. Flattening in the central part of the surface
density distribution of a globular cluster system is also seen in
other similar galaxies, such as M87
(McLaughlin \markcite{mcl95} 1995) and NGC 5128
(Harris \etal \markcite{har84} 1984). Our measurements
for the distribution of old clusters are in good agreement with those of
Kaisler \etal \markcite{kai96} (1996). However, they adopt a power law slope of -1.3,
based on the observational relation between galaxy magnitude and power
law slope of the surface density distribution of the globular cluster system
(Harris \etal \markcite{har86} \markcite{har93} 1986, 1993).
We note that our measured outer
slope is sensitive to the choice of break point
between the inner and outer regions; the data show a sign of steepening even
more in the outermost parts.
The spatial distribution of the clusters agrees with
the galaxy merger model of Ashman and Zepf \markcite{az92} (1992), which predicts that
the younger clusters will be peaked towards the center of the
galaxy, while the older clusters will be more flattened in their
distribution.

\subsection{Specific Frequency}

The specific frequency of globular clusters in a galaxy
is given by

$$S_N = N_T 10^{0.4(15+M_V)}$$

\noindent 
(Harris and van den Bergh \markcite{har81} 1981)
where $N_T$ is the total number of clusters and $M_V$ is the
absolute V magnitude of the galaxy. This is of interest in light of the
hypothesis that elliptical galaxies form via mergers of spirals
and the observation
that ellipticals have a higher specific frequency
of globular clusters than spirals.
Consequently, the merger hypothesis requires that globular clusters form
during mergers with
a higher specific frequency than that of the progenitor
population. Also, it requires that clusters form in numbers comparable
to those of the old cluster population (Zepf \& Ashman \markcite{az93} 1993).
To address these questions, we
calculate the specific frequency of the old population, the
current specific frequency of the merger related population, and the 
predicted specific frequency of the merger related population in 13 Gyr.

To calculate specific frequencies,
we need to estimate the total number of clusters associated
with each population.
These estimates come from
integrating the surface density profiles of the red and blue clusters
from the galaxy core to 100 kpc. This integration suggests that the total
number of blue clusters is five times the number of clusters that we observe.
For the red clusters, we correct this number
for the clusters fainter than our completeness cutoff
based on a
Gaussian fit to the part of the magnitude distribution which we have
observed.
For the blue clusters, we make no correction for missing faint clusters
and consider the resultant specific frequency to be a lower limit.
To make a rough prediction of the number of clusters
from the merger related population which 
survive to be 13 Gyr old, we assume
that the cluster luminosity function will come
to resemble the globular cluster luminosity function of the Galaxy after
4 magnitudes of fading and substantial cluster destruction.
We assume that no clusters brighter than the predicted
turnover of M$_B$ = -10.6 are destroyed,
and that the remaining clusters fainter than the turnover are equal in
number to those brighter than the turnover, forming a Gaussian luminosity
function. We note that this would require the
destruction of $\sim$ 90\% of the young clusters in NGC 1275 seen within our
field.
If the young cluster population is older than 10$^8$ yrs,
less cluster destruction would be required, and we would derive a higher
specific frequency. Of course, if some fraction of the massive clusters
are destroyed as well, we would derive a lower specific frequency.
Our corrections yield estimates of 12,700 clusters for the old population,
5700 clusters for the young population, and 550 clusters remaining from
the young cluster population after 13 Gyr.

We also need to estimate what fraction of the underlying stellar component
is associated with
each of the populations. NGC 1275 clearly has an excess blue population
over what is normally observed in central cluster galaxies (Romanishin \markcite{rom87} 1987).
It is unknown whether this blue population is related to star formation
in a recent merger or whether it is related to the presence of the cooling
flow centered on NGC 1275. Assuming the entire excess blue light comes
from a merger-related population, we derive a lower limit for the specific
frequency of globular clusters in such a population. Romanishin \markcite{rom87} (1987) 
estimates that 15\% of the total B light comes from an excess blue population,
and we adopt this contribution as an upper limit for the V band as well.
With a total brightness of V=11.88 (deVaucouleurs \etal \markcite{dev91} 1991), our
adopted $A_V=0.53$, and $H_0=75$ km/s/Mpc, we derive $M_V^{tot}=-22.89$.
Using a 15\% contribution for the merger-related population gives
$M_V^{young}=-20.83$ and $M_V^{old}=-22.71$. In order to estimate
the specific frequency of the merger-related population {\it as it would
appear in $\sim$ 13 Gyr}, we need to allow for a fading of the background
population, and we adopt 4 magnitudes of fading based on the population
synthesis models.
 
We combine the information on the total number of clusters with the
total V magnitudes for the populations
to calculate specific frequencies.
These estimates yield a $S_N\sim 10$ for the old population,
a $S_N\sim 27$ for the current young population (without
corrections for background fading or for cluster destruction), and
a $S_N\sim 102$ as an estimate of the specific frequency of the
young population
after 13 Gyr of cluster destruction and 4 magnitudes of
background fading.
Even without correction for clusters outside the WFPC2 field of view,
we would predict 150 clusters surviving for 13 Gyr
and a specific frequency of 28.
All of these
numbers, however, have significant uncertainties, as the corrections
for completeness, cluster destruction, and fading are very substantial
while at the same time being highly uncertain. From
shallower ground-based observations
over a larger field, Kaisler \etal \markcite{kai96} (1996) determine the S$_N$ for the old 
cluster population to be 4.3. The main sources of disagreement are that
they have adopted a steeper surface density profile and a different
total V magnitude.

The increase in the \it{overall}\rm \ specific frequency of
the galaxy from this event, however, is small, as the number
of these clusters predicted to survive
13 Gyr is small; only about 550 new clusters would be added
to a population of 12,700. Whether
the specific frequency increases significantly in a merger depends on the
mass and gas content of the merging bodies.
However, the numbers are intriguing because they suggest that,
even allowing for
substantial cluster destruction, the specific frequency of a merger-related
population in 13 Gyr will be much higher than that typically observed in 
spiral galaxies. While this particular merger event will
not have a large impact on the overall specific frequency
in NGC 1275, it does show that 
a merger population can have a large specific frequency, which is the
first requirement for increasing the overall specific frequency in a merger.
Consequently, the observation that ellipticals have
a higher specific frequency of globular clusters is not inconsistent with
the hypothesis that these galaxies have formed via the merging of spiral
systems.

Our calculation indicates that an estimate of the future specific
frequency is difficult to make, as several large corrections must
be applied. The most uncertain of these is probably the estimate of the
number of clusters which will be destroyed. Future work will need to
carefully consider the probability of cluster destruction as a function
of cluster mass and density if one wants to put serious constraints on
the merger hypothesis for ellipticals from observations of their
globular cluster systems.

\subsection{Sizes}

Holtzman \etal \markcite{hol96} (1996) estimated the sizes of marginally resolved clusters
in WFPC2 images using the difference between a 1 and a 2 pixel aperture 
magnitude. These measurements are compared with measurements of simulated
objects made by convolving a model PSF with a modified Hubble profile.
The comparison yields an estimate of the cluster core 
radius {\it if} the cluster shape is well modelled by a modified
Hubble profile.  Figure 10 shows measurements of $m_1-m_2$ in F702W for
the NGC 1275 objects; note that values are near zero because $m_1-m_2$
is defined differentially with respect to a model point source placed
at the same location on the frame as each cluster (see Holtzman \etal \markcite{hol96} 1996
for details). Positive values of $m_1-m_2$ indicate that the objects
are resolved, and core radii are estimated using Figure 4 of Holtzman \etal
\markcite{hol96} (1996).
For the brighter objects, we find that the objects are most likely resolved,
with estimated core radii of $\sim 0.01-0.05$ pixels, corresponding to $\sim$
0.15-0.75 pc using our adopted distance to NGC 1275.
Despite the uncertainties in these
measurements, it seems clear that these objects are compact, and much more
comparable to globular clusters than to open clusters.

\section{Conclusions}

From new deep observations of NGC 1275 with HST, we identify roughly 3000
objects which appear to be compact star clusters.
The color distribution of the cluster system is bimodal,
with a blue population which has (B-R)$_0$ $\sim$ 0.4
and a red population which has (B-R)$_0$ $\sim$ 1.3.
We suggest that the red objects are members of
the old globular cluster system and
that the blue objects are members of a young globular cluster system.
In an apparently dust-free region,
the spread in the colors of the blue clusters is small enough that it
is entirely attributable to scatter from errors in the
photometry, mostly due to errors in sky subtraction. This suggests that
the blue clusters are a single color, single age population. This
argues against the hypothesis that clusters have been forming continuously
from the cooling flow, and
supports the hypothesis that their formation  may have been triggered by
a previous merger.

The luminosity function continues to rise to the limit of our observations 
and is
inconsistent with a Gaussian globular cluster luminosity function
peaking near M$_V$ = -7.3 after correction for evolutionary
effects.  A similar luminosity function observed in NGC 4038/9 (the
Antennae) has been used by van den Bergh \markcite{vdb95} (1995) to argue that the
young clusters observed there are not true globular clusters.
However, the masses and sizes of the individual young clusters appear to be
comparable to those of globulars. We suggest that either the 
luminosity function evolves, with fainter clusters being preferentially
destroyed as time passes, or that the initial luminosity/mass function of the
young cluster system is different from that of typical old globular
cluster systems. 

If cluster destruction is responsible for the difference in luminosity
function, a destruction mechanism which preferentially destroys fainter,
lower mass clusters must be invoked. In order to create a turnover in the
luminosity function at the same mass as in old
cluster systems, the destruction of over 90\% of the clusters currently
seen in NGC 1275 would be required, given our assumption of 4 magnitudes
of fading.

Given these likely differences in the luminosity functions of
different globular cluster
systems, whether caused by evolutionary effects or by initial differences,
it is clear that some caution must be exercised in using the cluster
luminosity function as a distance indicator.

We have attempted to estimate the specific frequency of the young population
seen in NGC 1275 to determine whether it is likely that the overall
specific frequency as seen at a future time could be increased because of
the proposed merger-related formation of globulars. We note that the overall
specific frequency of NGC 1275 will only be slightly increased by the
proposed merger event which formed the blue clusters, due to the small
number of clusters expected to survive for a Hubble time. But we estimate
a large specific frequency of the merger related population, which suggests
that mergers are efficient in forming globular clusters.
However, corrections
for incompleteness at faint magnitudes, clusters residing outside the
current field of view, uncertainties in the mass and luminosity of
the stellar population formed during a merger event, and, especially,
corrections for evolutionary effects in both the background and cluster
populations are all significant effects for which we have only
relatively crude estimates.

\acknowledgements

This work was supported in part by NASA under contract
NAS7-918 to JPL and a grant to M. C. from the New Mexico Space
Grant Consortium.

\newpage

\begin{figure}
\caption[Carlson.fig1a.ps,Carlson.fig1b.ps]{a) PC image of
NGC 1275. b) WFPC2 image of NGC 1275.}
\end{figure}

\epsscale{0.6}

\begin{figure}
\caption[Carlson.fig2.ps]{Scatter in measured magnitudes of artificial
stars. 
The simulated stars were given a B-R of 0.7. Vertical dashed lines represent 
the input magnitudes for all artificial clusters of a given point type. Only 
clusters in the SW corner of the image with errors less than 0.15 mag in
B and R are shown, to simplify comparison to Figure 4b.}
\vskip 0.25in
\plotone{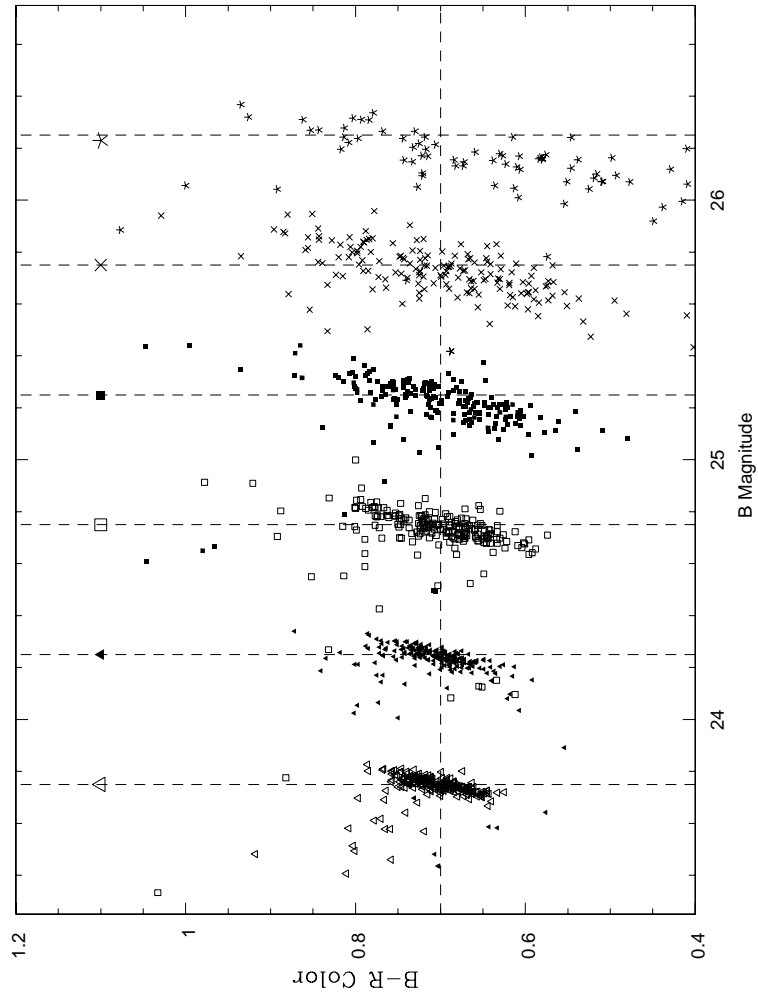}
\end{figure}

\begin{figure}
\caption[Carlson.fig3.ps]{Completeness as a function of B magnitude for the clusters
in the PC frame and in the whole image (weighted by the distribution of
clusters). Blue and red clusters are shown separately.}
\plotone{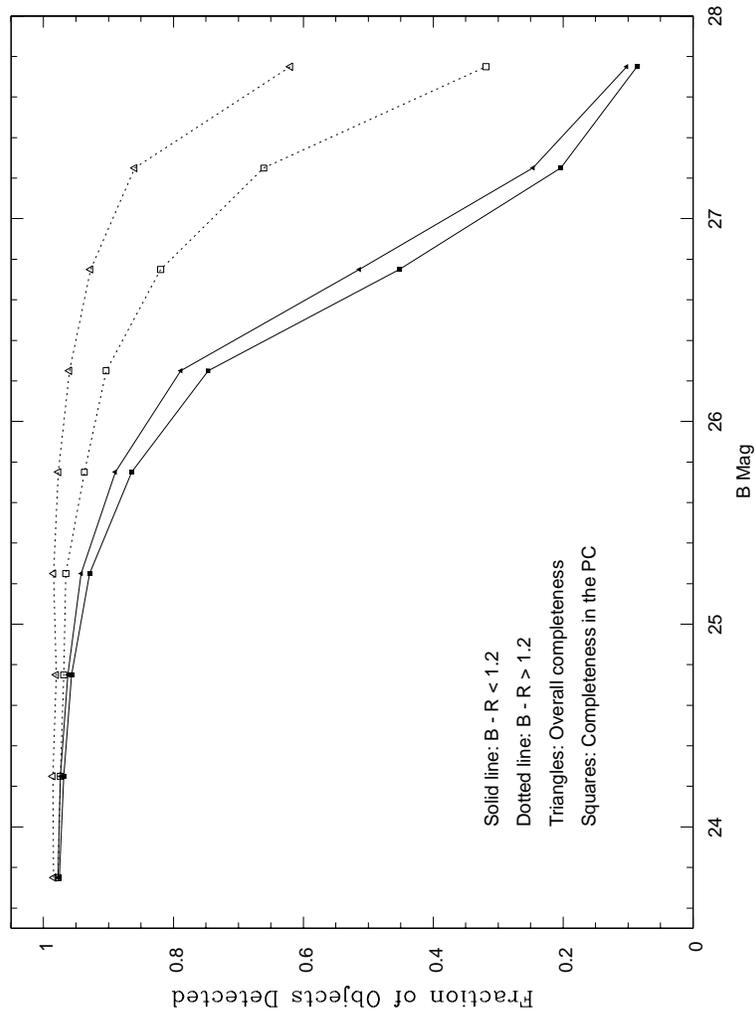}
\end{figure}

\epsscale{0.45}
\begin{figure}
\caption[Carlson.fig4a.ps,Carlson.fig4b.ps]{a) B-R color vs. B
magnitude for all clusters in the PC and WFs. 
b) B-R color vs. B
magnitude for clusters in the southwestern portion of the PC frame with
errors less than 0.15 magnitudes in both B and R.}
\vskip 0.5in
\plottwo{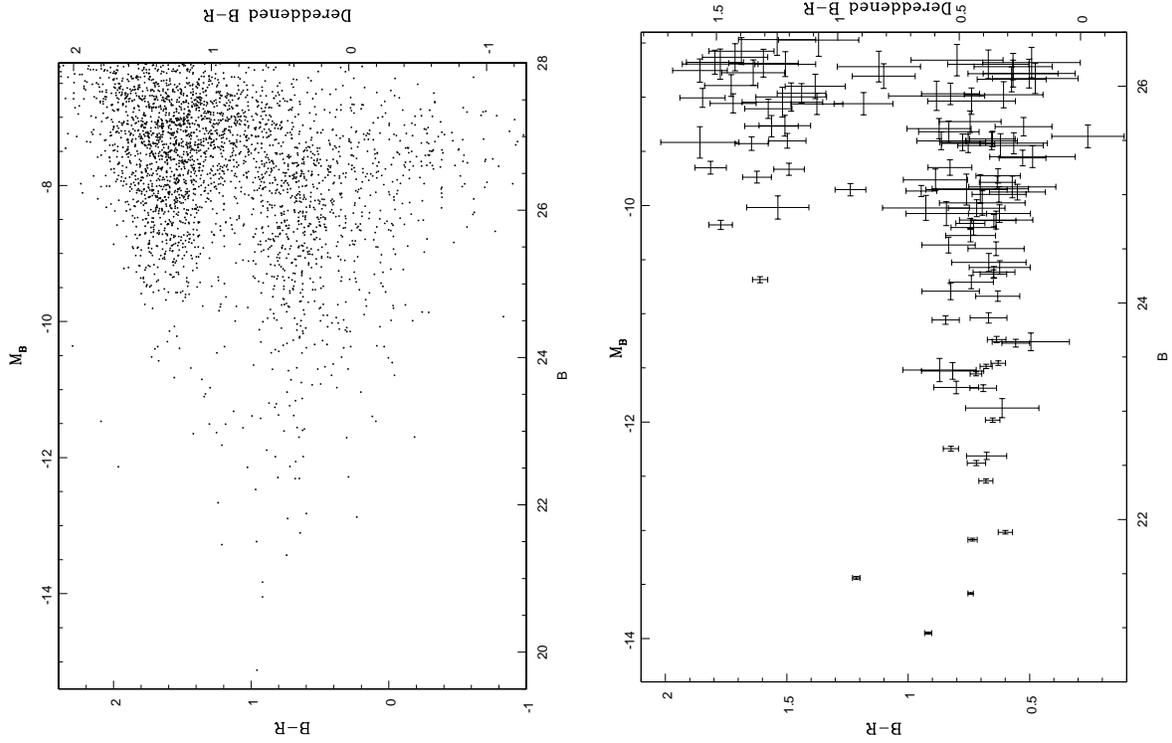}{Carlson.fig4b.epsi}
\end{figure}
\epsscale{0.6}

\begin{figure}
\caption[Carlson.fig5.ps]{Luminosity and
color evolution for a single burst, simple stellar
population from Charlot and Bruzual (1991). The
luminosity is scaled to a 1 M$_\odot$
object. Evolution is shown for 3 different metallicities.}
\plotone{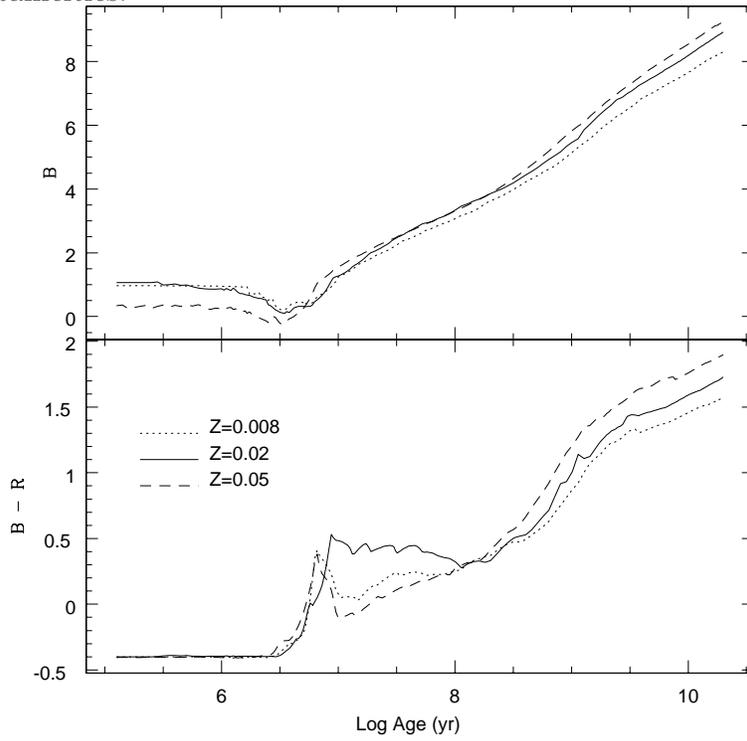}
\end{figure}

\begin{figure}
\caption[Carlson.fig6.ps]{B-V color vs. age for LMC clusters.}
\vskip 0.25in
\plotone{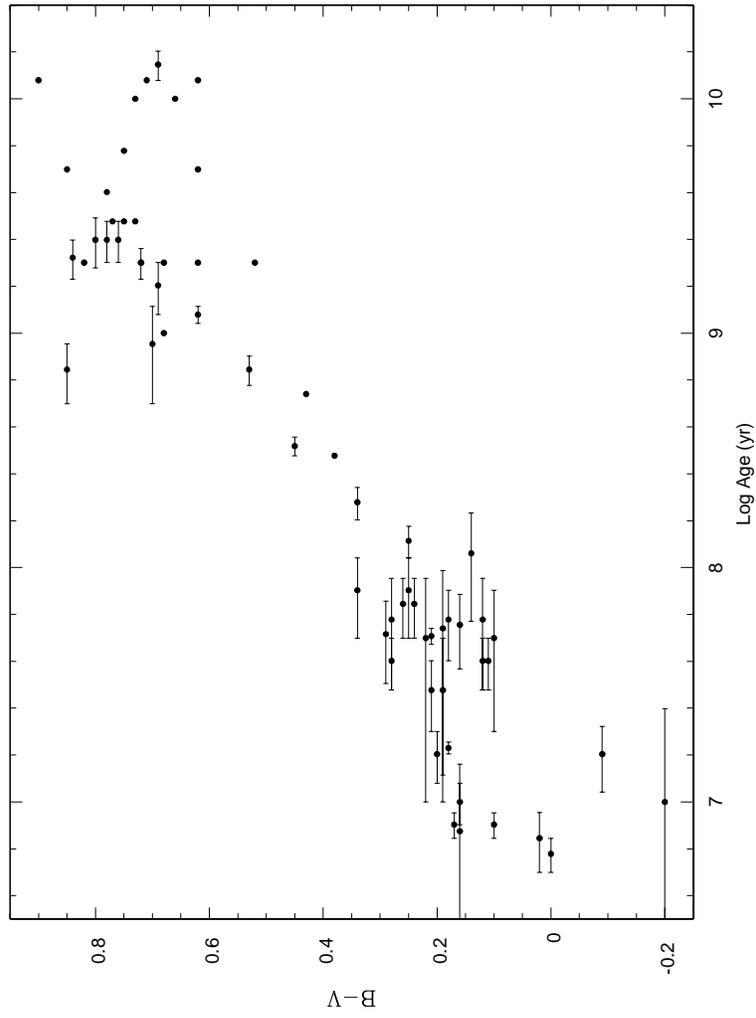}
\end{figure}

\begin{figure}
\caption[Carlson.fig7.ps]{Luminosity function for the clusters (solid) and
with correction for
completeness (dotted). a) The blue clusters
(B-R $<$ 1.1) b) The red clusters (B-R $>$ 1.1).}
\plotone{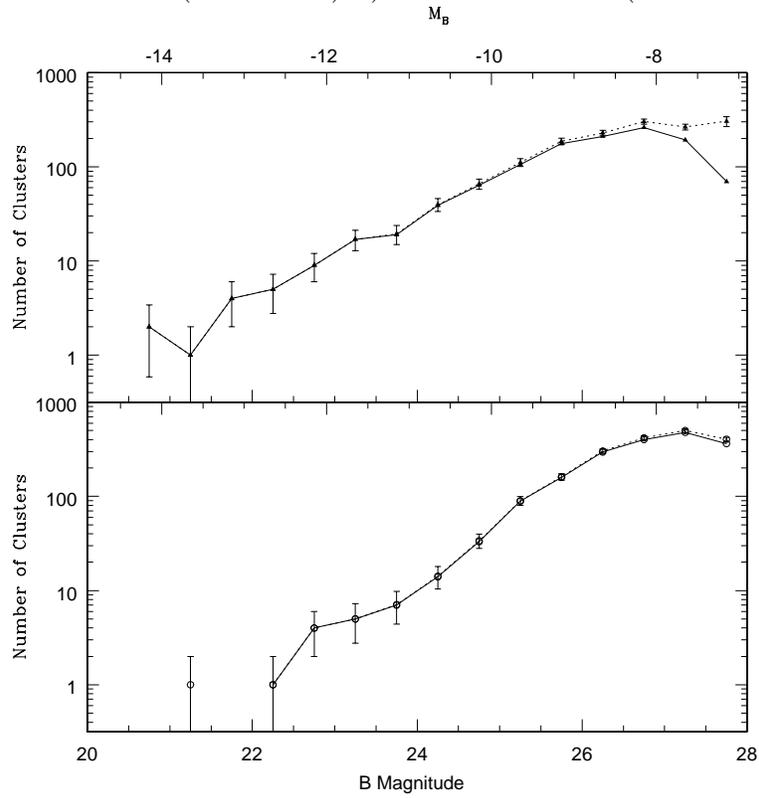}
\end{figure}

\begin{figure}
\caption[Carlson.fig8.ps]{Luminosity functions for the Milky Way and five
blue cluster galaxies. a) Milky Way globulars (Harris 1996) and
open clusters (van den Bergh \&
LaFontaine 1984),
b) NGC 1275, c) NGC 3921 (Schweizer \etal 1996), 
d) NGC 3597 (Holtzman \etal 1996),
e) NGC 4038/9 (Whitmore \& Schweizer 1995),
f) NGC 7252 (Whitmore \etal 1993) . 
Completeness limits are shown as dashed lines (where available).
The predicted turnover in the luminosity function if the population is identical
to Galactic globulars, but younger, is shown as a dotted line.}
\plotone{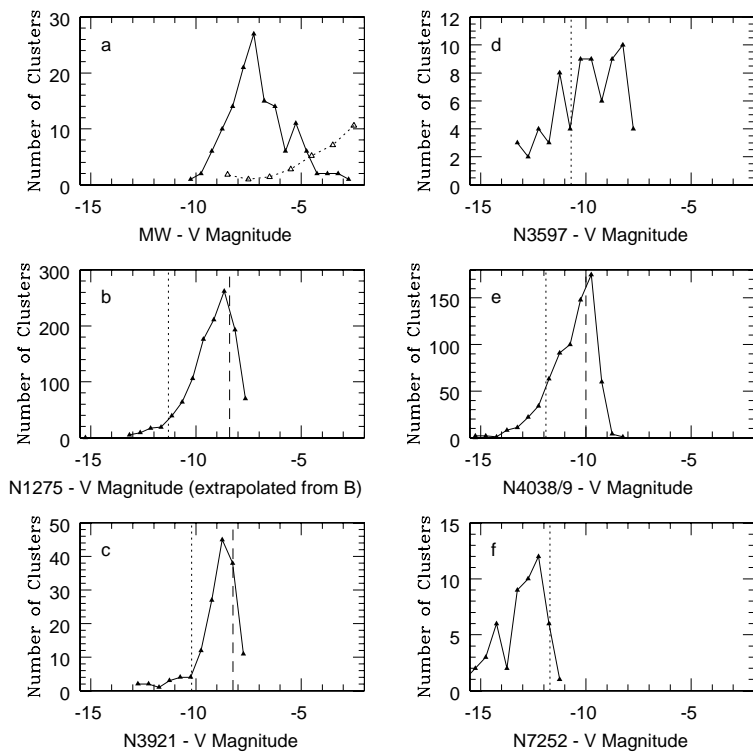}
\end{figure}

\begin{figure}
\caption[Carlson.fig9.ps]{The surface density profile
of the red clusters (B-R $>$ 1.1, solid
line) and the blue clusters (B-R $<$ 1.1, dotted line).}
\vskip 0.25in
\plotone{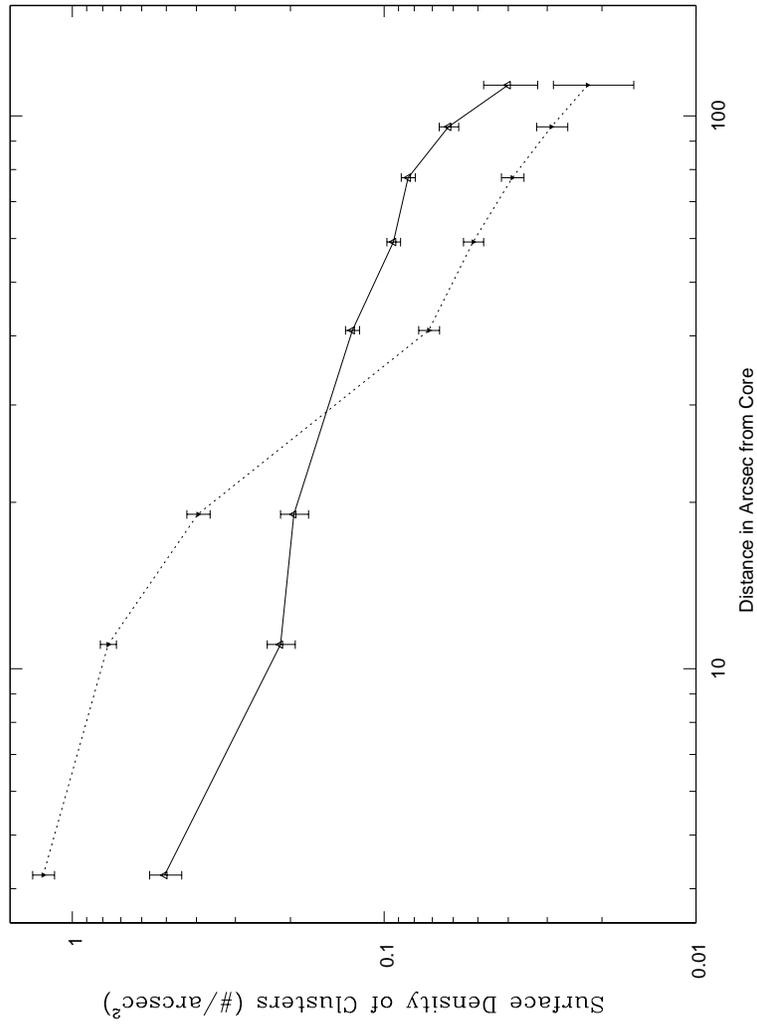}
\end{figure}

\begin{figure}
\caption[Carlson.fig10.ps]{Measurements of $m_1-m_2$ in F702W for all objects in the PC, where
$m_1-m_2$ is the difference between a one and two pixel aperture magnitude
relative to the same quantity measured on a model point source.
Positive values suggest that the objects are marginally resolved.}
\plotone{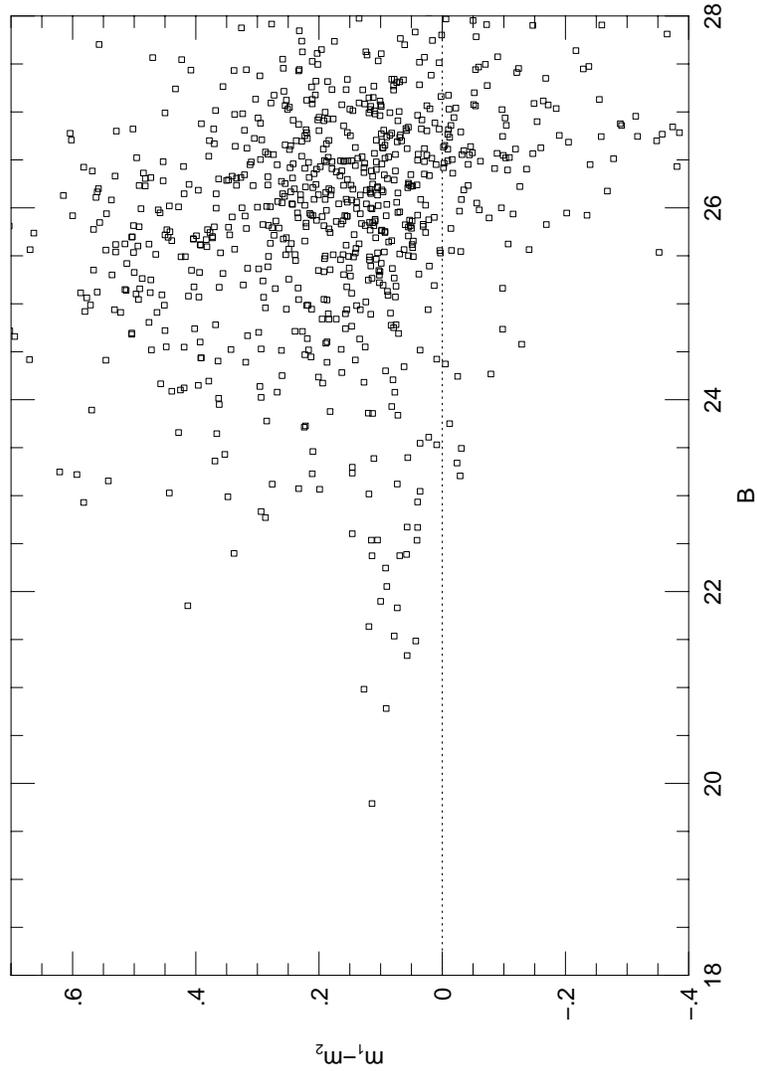}
\end{figure}

\end{document}